\documentclass[aps,pre,onecolumn,showpacs,showkeys,a4paper]{revtex4}
  \input epsf
\usepackage{graphicx}
\usepackage{amsmath}
\usepackage{amssymb}
\usepackage{amscd}

\def\*{******************************************}

\begin{document}
\title{A field-theoretic approach to nonequilibrium work identities}

\author{Kirone Mallick }
 \affiliation{Institut  de Physique Th\'eorique, Centre d'Etudes de Saclay,
 91191 Gif-sur-Yvette Cedex, France}
 \author{Moshe Moshe}
\affiliation{Department of Physics,
  Technion - Israel Institute of Technology, Haifa ~32000, Israel}
  \author{Henri Orland}
 \affiliation{Institut  de Physique Th\'eorique, Centre d'\'Etudes de Saclay,
 91191 Gif-sur-Yvette Cedex, France}



\date{\today}
\begin{abstract}
     We study   nonequilibrium work relations   for a
space-dependent field with stochastic dynamics (Model A).  Jarzynski's
equality  is obtained  through  symmetries of  the dynamical action in
the  path  integral representation.  We  derive a set of exact
identities that generalize the fluctuation-dissipation relations to
non-stationary and  far-from-equilibrium situations. These identities
are prone to experimental verification.  Furthermore, we show that a
well-studied invariance  of the Langevin equation  under
supersymmetry,  which is known to be  broken when the external
potential is time-dependent, can be  partially restored by adding to
the action  a term which is precisely Jarzynski's  work.  The work
identities can then be  retrieved as  consequences of the associated
Ward-Takahashi identities.
\end{abstract}
 \maketitle

 \section{Introduction}

  During the last decade, many  exact relations
 for non-equilibrium processes have been  derived. The
 Jarzynski equality is one of these remarkable results:
 \begin{equation}
  \langle {e}^{-\beta W}   \rangle  =  {e}^{-\beta \Delta F} \, .
 \label{eq:Jar1}
 \end{equation}
  This relation implies  that the statistical properties of the
  work performed on a system in contact with
 a heat reservoir at temperature $kT = \beta^{-1}$
 during a {\it  non-equilibrium}  process are
  related  to the free energy difference $\Delta F$
   between two  {\it equilibrium}  states of that system.
   This identity  was  derived originally using a Hamiltonian
 formulation \cite{jarzynskiPRL} and was  extended to
 systems obeying  a
 Langevin equation  or a discrete  Markov equation  \cite{jarzynskiPRE}.
  This result was generalized  by Crooks  \cite{crooks1,crooks2}, who showed
  that the identity~\eqref{eq:Jar1} results from a remarkable  relation
  between  the  probability $P_F(W)$  of performing  the  quantity  of work $W$
  in  a given (forward)  process  and the
   probability $P_{R}(-W)$  of performing $-W$
  in the reversed process, namely
  \begin{equation}
   \frac{ {\mathcal P}_F(W) }{ {\mathcal P}_R(-W)} = {e}^{ \beta (W - \Delta F)  }
 \, .
 \label{Crooksintro}
 \end{equation}

 Jarzynski and Crooks' identities   are  now   well established results
 (a  review of the state of the art
 can be found for example in  \cite{jarzynskiEPJ}).
   These relations have  been verified
  on exactly solvable models \cite{jarmazonka}
  and  by explicit  calculations in
 kinetic theory of gases \cite{lua,vandenbroeck}.
 These  equalities have
   also  been used  in   various
 single-molecule pulling experiments
 \cite{hummer,liphardt,ritort} to measure folding free energies
 (for a review of biophysical applications
  see e.g. \cite{ritortrev}),
 and have  been checked against analytical predictions
 on  mesoscopic mechanical devices
 such as a torsion pendulum \cite{douarche}.
  Experimental verifications  are delicate  to carry out  because
 the  mathematical validity of Jarzynski's theorem
    is insured by rare events that occur  with a  probability
  that typically decreases  exponentially with the system size
 \cite{jarzynskiJSM}.

      Another type of  physical systems where
  large fluctuations are expected  to occur
   are extended statistical  models in the vicinity of a phase transition.
  Such systems are often modeled by a continuous
  space-time description with a local
  coarse-grained order parameter, $\phi(x)$, which minimizes
  a  Ginzburg-Landau type free energy.
  The equilibrium properties  of these models
   have  been thoroughly studied,  in particular
   using  renormalization group techniques \cite{lubensky,ma,JZJ}.  Besides, in
   the vicinity of a critical point, the  dynamic properties
 also display anomalous behaviour.
  If one assumes that universality
 remains valid \cite{ma}, it is natural to construct and investigate the simplest
 dynamical models, with a given static behaviour, which  respect
 some physical constraints such as symmetries and conservation laws.
 The coarse-grained  order parameter  of a microscopic system
 is then  represented by a space and time dependent field  $\phi(x,t)$
 that  evolves according to an
 effective stochastic  differential equation.  Different possible  types of evolution
 equations have been  classified (see e.g. the review paper of Halperin and Hohenberg
 \cite{hohenberg}).

  In the present work, we derive  nonequilibrium
 work relations   for a field $\phi(x,t)$ that follows the simplest
 dynamics in  Halperin and Hohenberg's scheme:  Model A  dynamics
  describes  the kinetic Ising model with  non-conserved order
 parameter.
 We represent  the stochastic  evolution as  a path integral
 weighted  by a dynamical action  \cite{JZJ}.
  Our  method   is closely related
 to the one used in   \cite{abishek1,jarzynski4} to  study the case
 of  a Langevin equation for a 0-dimensional scalar coordinate
 that depends only on time. Adding a spatial dependence
  allows us  to use the powerful response-field formalism
 that was developed in \cite{msr}. The  work relations are then derived
   from elementary  invariance properties of  the path integral  measure
  under  changes of variables that affect simultaneously
  the original field $\phi(x,t)$  and its conjugate
 response-field $\bar\phi(x,t)$. From the work relations, we derive
  correlator identities that generalize
 the  equilibrium fluctuation-dissipation relations
 for situations that can be  arbitrarily far from equilibrium.

 One advantage of  introducing  space and time varying  fields
 is to extend the possible symmetries of the system and to consider transformations
 that can mix space and time.  In the present context,
  this  can be achieved
  by introducing  two conjugate  auxiliary Grassmann fields, $c(x,t)$
 and  $\bar{c}(x,t)$.
  The new dynamical action,
 which now depends on four fields $\phi, \bar\phi, c$ and $\bar{c}$,
 exhibits a larger invariance which is  a manifestation of a  hidden
 supersymmetric invariance  \cite{parisi,tsvelik}.
  This  property  is true at  thermodynamic equilibrium and it  is known
 that the equilibrium   fluctuation-dissipation theorem
  as well as  the  Onsager reciprocity relations can be derived
  from it   \cite{chaturvedi,gozzi,gozzi2} (see also \cite{JZJ}).
  In other words,  supersymmetry
  is  a  fundamental  invariance  property of the full dynamical action   that
  embodies the  principle of microscopic  reversibility.
 For a system out  of equilibrium (for example  a system subject to  a time-dependent
  external drive),   supersymmetric invariance is broken.
 This   leads   to a violation  of
 the  fluctuation-dissipation theorem or, equivalently,
  to the occurrence of corrective
 terms in the formulation of this theorem:  this fact was
 clearly recognized in \cite{trimper,zimmer}. Here, we remark  that  weighing
  the expectation values by the Jarzynski term $e^{-\beta W}$ amounts to modifying
 the dynamical action of the model and we show that the modified action
 exhibits an invariance under a  specified supersymmetric transformation.
 This  invariance manifests itself
 as correlators identities known as the the Ward-Takahashi identities
 (a field-theoretic counterpart to Noether's theorem). Finally,
  we prove that the  nonequilibrium work relation  can be deduced from
   the Ward-Takahashi identity that encodes   the  underlying supersymmetry. Therefore,
   supersymmetric invariance  of stochastic evolution equation
  is a fundamental property  that embodies   equilibrium relations (Onsager
 reciprocity, fluctuation-dissipation theorem)
  as well as nonequilibrium work identities.

 The outline of this work is as follows. In Section~\ref{sec:proofpathintegral},
 we define the model, use the response-field formalism to derive the work relations
 for a space-time dependent  field, and obtain  a  fluctuation-dissipation relation,
 valid far from equilibrium, which can be shown to be mathematically
  equivalent to the Jarzynski relation. In Section~\ref{sec:SUSY}, we
  use  the  formalism of   Grassmann   fields  for  the
  equilibrium case and define  precisely the various
  supersymmetric transformations that leave the dynamical action invariant.
   We then   show  that supersymmetry,
 which is broken when the system is
 out of equilibrium,  is restored by  modifying  suitably the  action
 and we prove that Jarzynski's equation  can be viewed  as a consequence
 of the  Ward-Takahashi identity  that encodes   the restored
 invariance. Concluding remarks are given in  Section~\ref{sec:conclusion}.
 Technical details are given in the appendices. In particular, the superfield
 formalism and
 the derivation of the  Ward-Takahashi identities are
  recalled   in   Appendices \ref{app:SUSYformalism} and
 \ref{demoward}.

 \section{Stochastic evolution of a scalar field}
 \label{sec:proofpathintegral}

       In this section,  we derive the field-theoretic version
 of the non-equilibrium work relations for a system that obeys
 a time-dependent Ginzburg-Landau equation. The dynamics considered
 will be purely relaxational and we shall focus on the most elementary
 case with no conservation laws,  described by Model A dynamics.
 We shall express the Probability Distribution Function of the field
 at a given time as a path-integral. The Jarzynski and Crooks relations
 will be obtained via this path integral formalism.

 \subsection{Model A dynamics}

  We   consider  a scalar  field  $\phi(x,t)$ that evolves
  in   a  d-dimensional space  according to  Model A dynamics
  \cite{hohenberg}, given  by the following
 stochastic   equation of motion:

\begin{equation}
    \frac{ \partial \phi}{\partial t}(x,t)
   =
    - \Gamma_0 \frac {\delta {\mathcal U}[\phi(x,t),t]  }{\delta  \phi(x,t)} +   \zeta(x,t)  \, ,
 \label{modelA}
\end{equation}
  where the  dynamics is
governed by the time-dependent
 potential
\begin{equation}
{\mathcal U}[\phi(x,t),t]
  = {\mathcal F}_{GL}[\phi(x,t)] - \int d^dx \ h(x,t) \phi(x,t) \, ,
\label{def:Ut} \end{equation}
   $h(x,t)$ being   an external  applied field.
  The time-independent part of the potential assumes
 the familiar Ginzburg-Landau form and is given by
 \begin{equation}
  {\mathcal F}_{GL}[\phi] =
  \int {\rm d}^d x \{ \frac{1}{2} r_0  \phi^2 + \frac{1}{2}
  |\nabla\phi|^2  + \frac{u_0}{4} \phi^4 \} \, .
 \label{formuleF}
\end{equation}
 The fluctuating driving field  $\zeta(x,t)$ is assumed to be  a Gaussian
 white noise of zero mean value and of  correlations given by
\begin{equation}
  \langle  \zeta(x,t) \zeta(x',t') \rangle = 2  \frac{\Gamma_0}{\beta}
  \delta(t -t')\delta^d(x -x') \, ,
 \label{noisecorrelation}
\end{equation}
 where   $ \beta =  (kT)^{-1}$ is   the inverse
  temperature.
 The fact that the auto-correlation
 of the  noise satisfies  Einstein's fluctuation-dissipation
 relation  ensures that  the dynamics is microscopically
 reversible and obeys detailed-balance \cite{vankampen}.
 The Langevin equation~\eqref{modelA} for
 model A thus reads
\begin{equation}
    \frac{ \partial \phi}{\partial t}(x,t)
      = - \Gamma_0 f(\phi(x,t),t) +    \zeta(x,t) \, ,
 \label{modelAbis}
\end{equation}
 where, for later convenience,   we have defined
\begin{equation}
 f(\phi(x,t),t) =  \frac {\delta {\mathcal U}[\phi(x,t),t]  }{\delta  \phi(x,t)} =
 -  \nabla^2 \phi(x,t) +  r_0  \phi(x,t) +  u_0\phi^3(x,t)- h(x,t) \, .
\label{eq:deff}
\end{equation}

 The  dynamics of  the order parameter can also
 be described in terms of   the  Probability
 Distribution Function  (PDF) ${\mathcal P}( \phi_1 |  \phi_0 )$  of  observing
 the field  $\phi(x,t)= \phi_1(x)$ at time $t=t_f$,  knowing that
  the initial field  is
 $\phi_0(x)$ at time $t=0$.  By definition, this PDF is given by
\begin{equation}
 {\mathcal P}( \phi_1 |  \phi_0 ) = \langle \,
   \delta  \left(\phi(x,t_f)-  \phi_1(x) \right) \,  \rangle  \, ,
\label{eq:defPDF}
\end{equation}
 where the expectation value is taken over all possible realizations
 of the noise  $\zeta(x,t)$  between the initial and the final times
 (the initial value of the field $\phi_0(x)$  being fixed).
 Substituting the Gaussian measure for the noise, this expression becomes
 \begin{equation}
 {\mathcal P}( \phi_1 |  \phi_0 ) =
  \int  \,   {\mathcal D}\zeta(x,t)
  ~{e}^{  -\frac{\beta}{4\Gamma_0}\int {\rm d}^d x  {\rm d}t \zeta^2}
    \delta \left(\phi(x,t_f)-  \phi_1(x) \right) \,
 \,\,\,\,  \hbox{ with } \,\,\,\, \phi(x,0)=\phi_0(x) \, .
    \label{eq:defP}
\end{equation}
 This expression is nothing but a formal path-integral solution
 of the functional
  Fokker-Planck equation associated with the  Langevin dynamics~(\ref{modelA}):
\begin{equation}
 \frac{\partial P}{\partial t} =\Gamma_0
 \int d^dx  \frac{\delta }{\delta \phi}
 \left(  f(\phi,t)  P +   \frac{1}{\beta} \frac{\delta  P }{\delta \phi}
   \right) \, .
\label{eq:FP}
\end{equation}
 When the external  field  is constant in  time $h(x,t) = h(x)$,
 this  Fokker-Planck equation has a stationary solution, which
 is the equilibrium Gibbs-Boltzmann  distribution:
\begin{eqnarray}
  { P_{eq}}[\phi] = \frac{e^{-\beta {\mathcal U}[\phi] }}{Z[\beta, h]}
 \, \,  \hbox{ with } \, \,
  Z[\beta, h] = \int {\mathcal D}\phi
\   e^{-\beta{\mathcal U}[\phi] }  \, .
 \label{mesureinv}
\end{eqnarray}
 Finally, we recall  that the equilibrium free-energy $F[\beta,h]$
 is defined by
 \begin{eqnarray}
F[\beta,h] = - \frac{1}{\beta} \log  Z[\beta, h]  \, .
\end{eqnarray}

 \subsection{Dynamic action for  the Probability Distribution}

  The  probability  ${\mathcal P}( \phi_1 |  \phi_0 )$ of
 observing
 the field  $\phi_1(x)$ at time $t_f$ starting from
 $\phi_0(x)$ at time $t=0$ is given by  equation~(\ref{eq:defP}).
 We now  rewrite the path integral in terms of the variable
 $\phi(x,t)$  using the  response-field formalism
  of Martin-Siggia-Rose, de Dominicis-Peliti  and Janssen~\cite{msr}.
  We start with   the  following identity (that can be found
  e.g. in \cite{JZJ})
  \begin{eqnarray}
    1 &=&  \int   {\mathcal D}\phi_1(x)  \int_{\phi(x,0)= \phi_0(x)}^{\phi(x,t_f)= \phi_1(x)}
   {\mathcal D}\phi(x,t) \,\,  \delta
   \left( \dot\phi(x,t) + \Gamma_0 \frac {\delta {\mathcal U}}{\delta  \phi}
    -\zeta(x,t) \right) |\det{\bf M} |  \nonumber  \\
  &=&   \int_{\phi(x,0)= \phi_0(x)}
  {\mathcal D}\phi(x,t) \,\,  \delta
   \left( \dot\phi(x,t) + \Gamma_0 \frac {\delta {\mathcal U}}{\delta  \phi}
    -\zeta(x,t) \right) |\det{\bf M} | \,  .
  \label{eq:Identite1}
  \end{eqnarray}
  Note that in the second equality, the configuration  $\phi(x,t_f)$  of the field
  at time  $t_f$, appears as an integration variable (i.e. $0 < t \le t_f$).
 The linear operator ${\bf M}$ is defined as
 \begin{eqnarray}
  {\bf M} =  \frac{\delta \zeta(x,t) }{\delta \phi(x,t)} =
   \frac{\partial}{\partial t} + \Gamma_0 \frac{\partial
   f(\phi(x,t),t)}{\partial \phi} \, .
    \label{eq:defM}
 \end{eqnarray}
 The determinant of this operator can be written as
\begin{equation}
 \det{\bf M}  = \exp\{ {\rm Tr}(\log {\bf M}) \} =
  e^{\frac{\Gamma_0}{2} \int d^dx dt  \frac{\delta^2{\mathcal U}}  {\delta \phi^2}} \, ,
\label{eq:detM}
\end{equation}
 where the last equation is  found by discretizing the operator ${\bf M}$
 and  using the  Stratonovich convention  \cite{JZJ,vankampen}.
  We substitute
 this expression in the identity \eqref{eq:Identite1}
  and introduce  the response field
  $\bar\phi(x,t)$ that allows us  to rewrite the functional
  Dirac distribution  $\delta()$  as
 an exponential. Thus,   we obtain
   \begin{eqnarray}
    1 =\int_{\phi(x,0)= \phi_0(x)}
   {\mathcal D}\phi(x,t) {\mathcal D}\bar\phi(x,t)
   |\det{\bf M} |
   ~e^{-\int d^dx dt \bar\phi \{\dot\phi +
   \Gamma_0 \frac {\delta {\mathcal U}}{\delta  \phi}
   -\zeta \}} \,   \,\,\,\,\, \hbox{ where }  \,\,\,\,\,  0 < t \le t_f \, .
  \label{eq:onedelta}
\end{eqnarray}
 We  now insert this   identity~(\ref{eq:onedelta})
 in equation~(\ref{eq:defP}),   perform
 the  Gaussian integral
 over  the  noise variable $\zeta(x,t)$ and substitute
 the expression~\eqref{eq:detM} for the Jacobian of ${\bf M}$.
 Finally,  the following expression for the
 PDF is obtained:
 \begin{equation}
  {\mathcal P}( \phi_1 |  \phi_0 ) =
 \int_{\phi(x,0)= \phi_0(x)}^{\phi(x,t_f)= \phi_1(x)}
  {\mathcal D}\phi {\mathcal D}\bar\phi \,
  {e}^{ - \int {\rm d}^d x  {\rm d}t
  \Sigma(\phi,\dot\phi,\bar\phi)} \, .
 \label{pathintforP}
\end{equation}
   The PDF is thus  expressed as a path-integral over the order parameter
  $\phi(x,t)$, with an effective dynamical action
$\Sigma$   given by
\begin{eqnarray}
 \Sigma(\phi,\dot\phi,\bar\phi)=  \Gamma_0  \bar\phi(
     \frac{\dot\phi}{\Gamma_0} +
 \frac{\delta {\mathcal U}}  {\delta  \phi}  -\frac{\bar\phi}{\beta})
  - \frac{ \Gamma_0 }{2}
 \frac{\delta^2{\mathcal U}}  {\delta \phi^2}  \, .
  \label{eq:action}
\end{eqnarray}
   The non-equilibrium identities will   arise  from invariance
 properties of the path-integral with action $\Sigma$.

 \subsection{Non-Equilibrium  Correlations  Identities}

  We   consider  the case where the  applied field varies
 with time according to a  well-defined   protocol: for  $t\le 0$,
 we have $h(x,0) = h_0(x)$ and the system is in
 its stationary state;   for $t>0$, the external field  varies
 with time and
  reaches its    final  value $h_f(x)$
  after a finite time $t_f$  and remains constant  for $t \ge t_f$.
  The  values of the potential
  ${\mathcal U}$ for $t \le  0 $ and $t \ge t_f$ are
 denoted by  ${\mathcal U}_0$ and ${\mathcal U}_1$, respectively.

 Let  ${\mathcal O}[\phi]$  be a functional
 that depends on  the values of the field $\phi(x,t)$ for
 $0 \le t \le t_f$. The average of ${\mathcal O}[\phi]$
 with respect to the stationary
 initial ensemble and the stochastic evolution  between times
 $0$
 and $t_f$ is given by the path integral
\begin{eqnarray}
   \langle  {\mathcal O}  \rangle  &=&
  \frac{1}{Z_0} \int {\mathcal D}\phi_0(x)  {\mathcal D}\phi_1(x)
  \rm{e}^{- \beta {\mathcal U}_0[\phi_0] } \,\,
   \int_{\phi(x,0)= \phi_0(x)}^{\phi(x,t_1)= \phi_1(x)}
  {\mathcal D}\phi {\mathcal D}\bar\phi \,
  {\rm e}^{ - \int {\rm d}^d x  {\rm d}t
  \Sigma(\phi,\dot\phi,\bar\phi)}
  {\mathcal O}[\phi]  \nonumber \\
  &=& \int  {\mathcal D}\phi_1(x)  \int {\mathcal D}\phi_0(x)
 \frac{{e}^{- \beta {\mathcal U}_0[\phi_0] }}{Z_0}
 \langle  \phi_1 | {\mathcal O} | \phi_0 \rangle
\label{meanvalueofO}
 \end{eqnarray}
   where we have defined
\begin{eqnarray}
 \langle  \phi_1 | {\mathcal O} | \phi_0 \rangle
 =  \int_{\phi(x,0)=
\phi_0(x)}^{\phi(x,t_f)= \phi_1(x)}
  {\mathcal D}\phi(x,t) {\mathcal D}\bar\phi(x,t) \,
  {e}^{ - \int {\rm d}^d x  {\rm d}t
  ~\Sigma(\phi,\dot\phi,\bar\phi)}
  ~{\mathcal O}[\phi] \ \ .
 \label{defI}
 \end{eqnarray}
 Under  a  change  of the integration variable $\bar\phi$
 in equation~(\ref{meanvalueofO}),
  \begin{eqnarray}
          \bar\phi(x,t) \rightarrow  -\bar\phi(x,t) +
  \beta \frac{\delta {\mathcal U}[\phi(x,t),t]}  {\delta  \phi(x,t)} \, ,
 \label{eq:shift}
\end{eqnarray}
the path integral measure is   invariant  but not the action  $\Sigma$
 which   varies as
   \begin{eqnarray}
          \Sigma(\phi,\dot\phi, \bar\phi)    \rightarrow
  \Sigma(\phi,-\dot\phi, \bar\phi) +
  \beta {\dot\phi}  \frac{\delta {\mathcal U} [\phi(x,t),t]}
    {\delta  \phi(x,t)} \, .
 \label{eq:variation}
\end{eqnarray}
Noticing that
  \begin{eqnarray}
 {\dot\phi}  \frac{\delta {\mathcal U} [\phi(x,t),t]} {\delta  \phi(x,t)}
  = \frac{d{\mathcal U [\phi(x,t),t]}}{ dt} - \frac{ \partial{\mathcal U} [\phi(x,t),t]}
 {\partial{t}}  \, ,
\end{eqnarray}
 we obtain
  \begin{eqnarray}
 \int  {\rm d}^d x  \int_{0}^{t_f}~{\rm d}t \  {\dot\phi}
\ \frac{\delta {\mathcal U} [\phi(x,t),t]}  {\delta  \phi(x,t)} =
 {\mathcal U}_1[\phi_1] - {\mathcal U}_0[\phi_0] -  {\mathcal W}_J[\phi] \,.
 \label{eq:variation2}
\end{eqnarray}
 The last term in this equation represents  Jarzynski's  work, defined by
  \begin{equation}
  {\mathcal W}_J[\phi] =  \int_{0}^{t_f}
   {\rm d}t \, \frac{ \partial{\mathcal U}}{\partial{t}}
 =   -  
   \int {\rm d}^d x \, {\rm d}t  \,\,  \dot{h}(x,t)\phi(x,t) \, ,
 \label{eq:defWJ}
\end{equation}
 the last equality being a consequence of   equation~(\ref{def:Ut}).
The change of sign of the time derivative $\dot\phi$
  in  equation~(\ref{eq:variation})  is  now compensated  by the
 change of   variables  in the  path integral
 \begin{eqnarray}
   \left(\phi(x,t), \bar\phi(x,t)\right)
  \rightarrow  \left(\phi(x,t_f -t), \bar\phi(x, t_f -t)\right) \, .
\label{eq:Tsym}
  \end{eqnarray}
  This time-reversal  transformation
  leaves the functional  measure invariant and restores $\Sigma$
 to its original form but with a {\it time-reversed}  protocol
 for the external applied field $h(x,t) \rightarrow h(x,t_f -t)$.
Performing the above change of variables
 (\ref{eq:shift}) and (\ref{eq:Tsym})
 in
  equation~(\ref{defI}) and using
   equations~(\ref{eq:variation})
  and~(\ref{eq:variation2}),
 we find,
  recalling that the work ${\mathcal W}_J$
 is odd under time-reversal,
\begin{eqnarray}
 \langle  \phi_1 | {\mathcal O}| \phi_0  \rangle  =
 {e}^{ \beta ({\mathcal U}_0[\phi_0]-
 {\mathcal U}_1[\phi_1])}
  \langle  \phi_1 | {e}^{-\beta {\mathcal W}_J}\hat{\mathcal O}| \phi_0  \rangle_R
  \,\,  \, .
 \label{Crooks0}
\end{eqnarray}
 On the right hand side, the
  subscript $R$ on the expectation value  denotes a
 time-reversed protocol. The  notation with a hat  $\hat{}$
 over an operator
 denotes the   time-reversed operator, more precisely:
 \begin{equation}
  \hat{\mathcal O}[\phi] =
  {\mathcal O}[\phi(x,t_f -t)] \, .
 \label{def:hatO}
 \end{equation}
 Inserting this
  identity in equations~(\ref{meanvalueofO} - \ref{defI}) allows us
  to derive the following  general  relation:
\begin{eqnarray}
   {  \langle  {\mathcal O}  \rangle}
  =  \frac{1}{Z_0} \int {\mathcal D}\phi_0(x) {\mathcal D}\phi_1(x)
 {e}^{ -\beta{\mathcal U}_1[\phi_1]}
 \langle   \phi_0 |  {e}^{-\beta {\mathcal W}_J}
    \hat{\mathcal O} | \phi_1  \rangle_R
 =   \frac{Z_1}{Z_0}
  {  \langle  \hat{\mathcal O}
  {e}^{-\beta {\mathcal W}_J} \rangle_R} = {e}^{-\beta \Delta F}
  {  \langle  \hat{\mathcal O}
  {e}^{-\beta {\mathcal W}_J} \rangle_R}    \,\,  \, ,
  \end{eqnarray}
  where $\Delta F$ is the free energy difference
  between the final and the initial states.
   Finally,   redefining    ${\mathcal O}$
 as   ${\mathcal O} {e}^{- \beta W_J}$,  we deduce   that
 \begin{equation}
    {\langle  {\mathcal O} {e}^{-\beta {\mathcal W}_J}   \rangle}
 = {e}^{-\beta \Delta F}   {  \langle  \hat{\mathcal O}  \rangle_R} \, .
 \label{OhatO}
 \end{equation}
  When   ${\mathcal O} = 1$, we obtain
 Jarzynski's  theorem
\begin{equation}
  \langle {e}^{- \beta W_J}   \rangle  =  {e}^{-\beta \Delta F}   \, .
 \label{eq:Jarzynski}
\end{equation}
   Taking  ${\mathcal O}  =  {e}^{(\beta-\lambda){\mathcal W}_J}$,
 where $\lambda$ is an arbitrary real parameter,
  we derive   the following symmetry  property
\begin{equation}
       \langle {e}^{- \lambda W_J}   \rangle  =  {e}^{-\beta \Delta F}
      \langle {e}^{(\lambda -\beta) W_J}   \rangle_R  \, .
      \label{eq:Laplace}
\end{equation}
 The  Laplace transform of this equation
 leads  to  Crooks relation~\eqref{Crooksintro}  in its usual form
  \cite{crooks1,crooks2}:
\begin{equation}
   \frac{ {\mathcal P}_F(W) }{ {\mathcal P}_R(-W)} = {e}^{ \beta (W - \Delta F)  }
 \label{eq:Crooks}
\end{equation}
 where  ${\mathcal P}_F$ and  ${\mathcal P}_R$ represent
 the   probability distribution functions  of the work
 for the forward  and the  reverse processes, respectively.
 We emphasize that the  proof
  of Crooks and Jarzynski  identities  is  based  on  invariance
 properties of the path integral
  and does not involve any  a priori  thermodynamic definition of heat
 and work. The  expression~(\ref{eq:defWJ}) for  the
  Jarzynski work
  appears here  as a natural outcome of  this  invariance.
  It is  important to notice that
 time-reversal is crucial to obtain the general identity~\eqref{OhatO}
 and Crooks' theorem. However, it is known that
 Jarzynski's identity can be proved without assuming  time-reversal invariance
 \cite{hummer}.

\hfill\break
  We emphasize that   Jarzynski's  identity is valid only under
carefully defined  boundary conditions:
 (i) the system is  at thermal equilibrium at time $t=0$;
 (ii) During the finite time interval $0 \le t \le t_f,$   the  system
 is subject to  an external protocol
   that drives it away from equilibrium;
 (iii) After  the finite  time $t_f$, all time-dependent parameters are
   frozen: these fixed parameters define  a new  state
 of thermodynamic  equilibrium
 towards  which  the system  relaxes  after an infinite amount of time.
 According to this scheme,
   all   path integrals  must  range over
  the finite
 interval of time  $0 \le t \le t_f$ and  the expectation value of the  operator
 ${\mathcal O}$, defined  in \eqref{meanvalueofO},
  has to be  taken  with respect to
  the Boltzmann distribution at  $t =0$  and the uniform distribution
 at the final time $t_f$.  However,   these  stringent
  boundary conditions necessary for  Jarzynski's  identity to be valid, allow
 us  to embed  naturally all  the  path integrals over the
 infinite range of time  $-\infty < t < +\infty$
 by using  the following properties of the
   probability distribution:
 \begin{eqnarray}
\frac{1}{Z_0}
  {e}^{- \beta {\mathcal U}_0[\phi_0]}=
      \lim_{\tau\to -\infty} P(\phi_0  | \phi_{\tau}) \, ,
 \label{ergodicity} \\
       1 =  \int {\mathcal D}\phi(x,\tau)  P(\phi_\tau  | \phi_{t_f})
 \,\,\,\, \hbox{ for  any } \,\,\,\,  \tau  > t_f
 \, .  \label{normalization}
 \end{eqnarray}
 The first property assumes ergodicity (i.e.
 the  Gibbs-Boltzmann distribution is reached at time $t=0$
 by starting from any initial condition at $t = -\infty$).
The second  equality simply results from
 normalization. In terms of  path integrals, the first  expression becomes
\begin{equation}
  \frac{1}{Z_0}
  \rm{e}^{- \beta {\mathcal U}_0[\phi_0] } =
\int_{ \phi(x,-\infty)=  \phi_{-\infty}(x)}^{\phi(x,0)= \phi_0(x)}
  {\mathcal D}\phi(x,\tau) {\mathcal D}\bar\phi(x,\tau)
   \,\,
  \rm{e}^{ - \int {\rm d}^d x  {\rm d}t
    { \Sigma}(\phi,\dot\phi,\bar{\phi})}  \,
 \,\,\,\, \hbox{ for  } \,\,\,\,
 -\infty < \tau < 0 \, ,
\label{eq:extmoinsinfty}
 \end{equation}
 where the condition at $t = -\infty$ is taken to be  an  arbitrary value
 $\phi_{-\infty}$ (or more generally   a  distribution of
 values, normalized to 1).  Similarly, equation~\eqref{normalization}
 is rewritten as
 \begin{equation}
        1 =    \int_{\phi(x,t_1) = \phi_1(x)}
 {\mathcal D}\phi(x,\tau) {\mathcal D}\bar\phi(x,\tau)
   {\rm e}^{ - \int {\rm d}^d x  {\rm d}t
  {\Sigma}(\phi,\dot\phi,\bar\phi)}  \,\,\,\, \hbox{ where  } \,\,\,\,
  t_f  < \tau < \infty \, .
\label{eq:extplusinfty}
\end{equation}
 We   now consider an operator   ${\mathcal O}[\phi]$ that
  differs from a constant  only for $0 \le t \le  t_f $
  (i.e.  the operator ${\mathcal O}[\phi]$
 depends on the values taken by $\phi$ only over the  finite range of time $0 \le t \le t_f$).
 Using
  the relations~\eqref{eq:extmoinsinfty}  and ~\eqref{eq:extplusinfty},
  the  expectation value of  ${\mathcal O}[\phi]$,
  defined in  equation~(\ref{meanvalueofO}),
 can be  expressed  as
\begin{eqnarray}
    \langle  {\mathcal O}  \rangle  =
   \int {\mathcal D}\phi(x,\tau) {\mathcal D}\bar\phi(x,\tau)
  ~{e}^{ - \int {\rm d}^d x  {\rm d}t
  { \Sigma}(\phi,\dot\phi,\bar\phi)}
  ~~{\mathcal O}[\phi] \, \,\,\,\, \hbox{ for   } \,\,\,\,
  -\infty  < \tau < \infty \, ,
  \label{eq:pathint3.0}
\end{eqnarray}
 and  where  the space-time  fields  $\phi(x,t)$,  $\bar\phi(x,t)$  are
  integrated over an infinite range of time
 and over the whole space.  The
  only restriction   on this path integral is that the initial
 condition at  $ t = -\infty$ is fixed (or more generally,
 the values at $ t = -\infty$  are sampled  from  a   normalized distribution).
  We note  that  the Jarzynski term
  ${e}^{-\beta {\mathcal W}_J}$ is  equal to 1 outside the interval
 $0 \le t \le t_f$  and therefore   we can also write
\begin{equation}
    \langle   {e}^{-\beta {\mathcal W}_J}  \rangle  =
    \int {\mathcal D}\phi {\mathcal D}\bar\phi\,\,
  {\rm e}^{ - \int {\rm d}^d x  {\rm d}t
  { \Sigma}(\phi,\dot\phi,\bar\phi)} {e}^{-\beta {\mathcal W}_J}     \, .
  \label{eq:pathint3bis.0}
\end{equation}
  We   emphasize  that   the   boundary conditions
 at finite time,
 as well as the average over
 the Boltzmann factor,   have   been eliminated and all  path integrals
 are now evaluated over the full time line $-\infty < \tau < +\infty$.

\subsection{A  Non-Equilibrium  Fluctuation-Dissipation Relation}
\label{sec:FDRnoneq}

  The identity~(\ref{OhatO}), which is at the core of the
  work fluctuation relations, is valid for any choice of  the
  external field protocol. The free energy variation is  a function  only
  of the extremal values of the applied field  at $t_0 =0$ and $t = t_f$
  and is independent of  the values  at intermediate
  times.   Therefore,  performing
  functional derivatives  of the Jarzynski identity~(\ref{eq:Jarzynski})
  with respect to  $h(x,t)$ at an intermediate time $ t_0 < t < t_f,$
  and at position $x$,  results in new  identities. For example, we have
\begin{equation}
     \frac{1}{\Gamma_0}
 \frac {\delta  \langle    {e}^{-\beta W_J}  \rangle}
 {\delta h(x,t)}
 =  \langle \big( \bar\phi(x,t) - \frac{\beta}{\Gamma_0}
  \dot\phi(x,t) \big) {e}^{-\beta W_J}  \rangle
   =  0 \, .
 \label{eq:deriveejarz}
\end{equation}
  More generally, the $n$-th functional derivative
  of equation~(\ref{eq:Jarzynski})
  at intermediate times $t_1,\ldots t_n$ and positions
   $x_1,\ldots x_n$, gives the identity
\begin{equation}
    \langle    {e}^{-\beta W_J} \prod_{i=1}^n
  \big(\bar\phi(x_i,t_i) - \frac{\beta}{\Gamma_0}
  \dot\phi(x_i,t_i) \big)   \rangle
   =  0  \, .
 \label{eq:deriveeniemejarz}
\end{equation}
 Similarly, the  functional derivative  of equation~(\ref{OhatO}) leads to
\begin{equation}
   \langle (\bar\phi(x,t) - \frac{\beta}{\Gamma_0}
 \dot\phi(x,t)) {\mathcal O}  {e}^{- \beta W_J}  \rangle =
    {e}^{-\beta \Delta F} \langle{\hat{\bar \phi} (x,t)}
  {\hat {\mathcal O}}  \rangle_R  \, .
 \label{eq:derivee1}
\end{equation}
  In particular, equation~(\ref{eq:deriveejarz})~follows by choosing
  ${\mathcal O}=\hat{\mathcal O}=\hat{\bf 1}$ and taking into account
  that
  $\langle {\bar\phi}\rangle=0$. [Indeed,
 if we take  the functional derivative
 of equation~({\ref{meanvalueofO}}) with respect to $h(x,t)$
 for  ${\mathcal O}={\bf 1}$,
  we obtain  $\langle \bar\phi \rangle = \langle \frac{\dot\phi}{\Gamma_0} +
 \frac{\delta {\mathcal U}}  {\delta  \phi} \rangle = 0.$]

 For  the special  case ${\mathcal O}[\phi] =
 \phi(x',t')$, equation~\eqref{eq:derivee1} leads to:
 \begin{eqnarray}
 \label{eq:GenFDT0}
 \langle  \bar\phi(x,t) \phi(x',t') \rangle  - \frac{\beta}{\Gamma_0}
\langle  \dot\phi(x,t)   \phi(x',t')  {e}^{- \beta W_J}  \rangle =
    {e}^{-\beta \Delta F} \langle{\hat{\bar \phi} (x,t)}
  {\hat \phi(x',t')}  \rangle_R  \, ,
\end{eqnarray}
 where $\hat{\bar \phi}$ is obtained from ${\bar \phi}$ by time reversal
 as defined in equation~\eqref{def:hatO}.
The terms proportional to the response field ${\bar \phi}$
  in the correlators can be generated as follows: we consider a small
 perturbation  $h_1(x,t)$  that drives
  the system out of the  fixed protocol
  $h(x,t)$. However, we keep  the definition of the Jarzynski work
  unchanged so that  the perturbing field
  $h_1(x,t)$ is not included in $W_J$. The field  $h_1$   couples
 to ${\bar \phi}$ in the action $\Sigma$:  therefore performing
 functional derivatives with respect to $h_1$ amounts to  inserting
 the field  ${\bar \phi}$ inside  correlation functions.
  The previous equation can thus  be rewritten as
 \begin{eqnarray}
\frac{\beta}{\Gamma_0}
 \langle \dot\phi(x,t))  \phi(x',t')   {e}^{-\beta W_J} \rangle =
 \frac {\delta \langle \phi(x',t') {e}^{-\beta W_J} \rangle }
 {\delta h_1 (x,t)} \Big |_{h_1=0}  -  {e}^{-\beta \Delta F} \frac {\delta  \langle \hat \phi(x',t') \rangle_R}{\delta \hat h_1 (x,t)}\Big |_{h_1=0}  \,  .
\label{eq:GenFDT}
\end{eqnarray}
We emphasize that
  $W_J$  has to be measured for the fixed protocol $h(x,t)$.
 In this form, equation~\eqref{eq:GenFDT}  appears as
  an exact  generalization of the fluctuation-dissipation relation (FDR).
 The equilibrium  FDR \cite{callen, kubo,agarwal}
   is retrieved  by setting $W_J$ and $\Delta F$ to 0.
 This indeed corresponds to a system prepared in an equilibrium state,
 which is not subject to any macroscopic protocol (i.e. $h(x,t) =0$)
 but which is   driven slightly out of equilibrium by
  the small perturbation $h_1(x,t)$:
 \begin{equation}
   \frac{\beta}{\Gamma_0}  \langle
  \phi(x',t') \dot\phi(x,t)   \rangle  =
  \langle \phi(x',t') \bar\phi(x,t) \rangle -
   \langle \bar\phi(x',t') \phi(x,t) \rangle
  = \frac {\delta \langle \phi(x',t') \rangle }
 {\delta h_1 (x,t)} \Big |_{h_1=0}  -
 \frac {\delta  \langle \hat \phi(x',t') \rangle_R}
 {\delta \hat h_1 (x,t)}\Big |_{h_1=0}  \, .
    \label{eq:FDT}
 \end{equation}
 The fact that
 the equilibrium fluctuation-dissipation relation  can be deduced by diffentiation
 from Jarzynski's identity (or equivalently from Crooks' relations)
 has been understood by various authors
  (see in particular the works of R. Chetrite et al.
  \cite{chetrite1,chetrite2,chetritethese,chetrite}).
 This technique can also be used to find analogs of the
 FDR  at higher orders  \cite{gaspard}.
 Generalizations to
  non-equilibrium stationary states (NESS)
  have been also proposed \cite{chetrite2,seifert1,seifert2,maes},
  e.g.,   starting from
   the Hatano-Sasa relations which are the counterpart of
  Jarzynski's identity for a NESS \cite{joanny}.  We emphasize that
  the  identity obtained
  in equation~\eqref{eq:GenFDT} belongs to a  different class. We do  not
  consider  a linear perturbation near  a  state of thermodynamic equilibrium,
  or  near  a  NESS. Rather, we  first apply, as in Jarzynski's scheme,
  a  protocol to a system  initially in   thermodynamic equilibrium
  (that  can be  driven
  as far from  equilibrium as wished) and then,
   we apply  linear  perturbations  around this fixed  protocol: this leads
   to  a   new fluctuation-dissipation theorem  that relates
  out of equilibrium  and  nonstationary
  response functions to  nonequilibrium and nonstationary
  correlation functions.
  The insertion of the  Jarzynski factor ${e}^{-\beta W_J}$
   inside the correlators leads to formulae which are
     valid far from equilibrium and  look
   very similar to   equilibrium relations.
  The relation~\eqref{eq:GenFDT}   could be verified
   in single molecule pulling experiments where  the protocol corresponds
 to the pulling force $F(t)$ and  $\phi$ does not depend on space.
Then, all the quantities  that appear  in this relation  are susceptible to
  experimental measurements    by adding a small perturbation
  $\delta F(t)$ to the fixed  protocol  $F(t)$.

  The correlator identity~(\ref{eq:deriveejarz}) was  obtained
 as a consequence   of   Jarzynski's  equality~(\ref{eq:Jarzynski})
 by taking its  first derivative.  Conversely,  we show in
 Appendix~\ref{app:converse} that
  equation~(\ref{eq:deriveejarz})    implies
   Jarzynski's equality~(\ref{eq:Jarzynski}) and is therefore
 equivalent to it. This  converse  property will be used  in the next section
 to show
  that  the work relation can be extracted from a hidden supersymmetric
 invariance of the dynamical action.

 \section{Supersymmetry and Nonequilibrium Work Relations}
 \label{sec:SUSY}

 Identities between correlators such as
   equations~(\ref{eq:deriveejarz})-(\ref{eq:derivee1})   suggest  the existence
 of an underlying continuous  symmetry of the  system. Indeed, it was recognized
 in the late seventies that the Langevin equation possesses a hidden invariance
 under supersymmetric transformations. This property was first
  discussed in the context
 of dimensional reduction \cite{parisi} and then used to derive convenient
 forms for diagrammatic expansion techniques \cite{tsvelik} that were used to
 study critical dynamics of relaxational models \cite{zinn,zhang,niel}.
 Supersymmetry was also used to study mesoscopic quantum systems
 (see e.g. in \cite{efetov,zirnbauer}).
 This supersymmetry became an efficient tool to study
  \cite{cooper, bernstein,comtet,keung}  the  properties
 of Fokker-Planck and associated Schr\"odinger  operators
 (see e.g. \cite{risken,junker}).
 It was also realized that the equilibrium fluctuation-dissipation  relation
 and the Onsager reciprocity relations could be derived from this  invariance
 \cite{chaturvedi,gozzi,gozzi2,zhang}. Conversely, nonequilibrium situations
  were found  to correspond to supersymmetry breaking and corrections
 to the classic equilibrium relations could formally
 be calculated \cite{trimper,zimmer}. In this section, we extend
  this  investigation  further by showing that although
  supersymmetry is broken under  nonequilibrium situations,
  it is partially recovered  by
  adding  to  the  dynamical action a term, which precisely corresponds
 to Jarzynski's work.  This restored  invariance leads to
    Ward-Takahashi identities  amongst
 correlation functions. Jarzynski's relation results from these
 identities. Thus, it is emphasized here that supersymmetry is
 not imposed but it is rather a result, namely an underlining
 symmetry of this system.

\subsection{Supersymmetric Invariance for the time-independent  Model A}

   First, we  consider the case where the external field
 does not depend on time:  the Langevin equation~\eqref{modelA} has
 then a well-known  supersymmetric invariance.  We shall review
 the formalism that allows to make this  invariance explicit, write
 the Ward-Takahashi and derive the equilibrium
fluctuation-dissipation  relation
 following \cite{chaturvedi}.

\subsubsection{The Supersymmetric Action  and its Invariance Properties}

  To uncover this  hidden symmetry,  we introduce  in addition to
 the original field $\phi(x,t)$ and  the response field
 $\bar\phi(x,t)$,   two  auxiliary
 anti-commuting  Grassmann  fields $c(x,t)$ and $\bar{c}(x,t)$
  that allow us to express  the Jacobian of ${\bf M}$, defined in
 equation~(\ref{eq:defM}),   as a functional
 integral \cite{chaturvedi,gozzi,JZJ}. These fields
   $c(x,t)$ and $\bar c(x,t)$
  can be viewed as hidden classical
  fermionic fields  that ensure the volume conservation
 constraints: they allow to enforce this conservation property
 at a  dynamical level. Inserting the following  identity \cite{JZJ}
  \begin{eqnarray}
\det{\bf M} = \int  {\mathcal D}c {\mathcal D}\bar{c}
 \rm{e}^{ c {\bf M} \bar{c}} =
\int  {\mathcal D}c {\mathcal D}\bar{c}
 \rm{e}^{ c(\frac{\partial}{\partial t} + \Gamma_0
 \frac{\delta^2{\mathcal U}}  {\delta \phi^2}) \bar{c}} \, ,
   \end{eqnarray}
 we observe that the PDF can be rewritten as
 \begin{eqnarray}
     {\mathcal P}( \phi_1 |  \phi_0 )  =
 \int_{\phi(x,0)= \phi_0(x)}^{\phi(x,t_1)= \phi_1(x)}
  {\mathcal D}\phi {\mathcal D}\bar\phi
  {\mathcal D}c {\mathcal D}\bar{c} \,\,
  \rm{e}^{ - \int {\rm d}^d x  {\rm d}t
  {\bf \Sigma}(\phi,\bar\phi,c,\bar{c})}  \,  ,
 \label{PathInt}
\end{eqnarray}
 where the   effective Lagrangian ${\bf \Sigma}$, which is now a function of the
 Grassmann variables as well,  is   given by
\begin{eqnarray}
 {\bf \Sigma} (\phi,\bar\phi,c,\bar{c})=  \Gamma_0  \bar\phi(
     \frac{\dot\phi}{\Gamma_0} +
 \frac{\delta {\mathcal U}}  {\delta  \phi}  -\frac{\bar\phi}{\beta})
  - c(\frac{\partial}{\partial t} + \Gamma_0
 \frac{\delta^2{\mathcal U}}  {\delta \phi^2}) \bar{c}\   \,\, .
 \label{eq:susyaction1}
\end{eqnarray}

  The action ${\bf \Sigma}$ exhibits
  two  important  invariances under infinitesimal  transformations that mix
 ordinary fields  with  Grassmann   fields. We  shall now  describe them
  by specifying  how each field varies under these transformations.
 In the Appendix~\ref{app:SUSYformalism}, we shall use  a more elegant
 presentation in which  the four fields $(\phi,\bar\phi,c,\bar{c})$
  appear to be  the components of a unique superfield ${\bf \Phi}$; also,
  in  this language,
 the dynamical   action  ${\bf \Sigma}$ will take a more  compact  form and
 the infinitesimal  transformations that leave it invariant
  will have a simple interpretation.

\hfill\break

  $\bullet$ {\bf Invariance under  BRST1  Transformation:}
 Consider $\epsilon$ to be  a time-independent  infinitesimal Grassmann field.
   We consider the following  transformation (that we  call   BRST1):
\begin{eqnarray}
   \delta \phi(x,t) &=&  - \bar{c}(x,t) \epsilon \, ,  {\hskip 1cm}
 \delta \bar{c}(x,t)  =   0   \, ,   \nonumber    \\
   \delta  c(x,t) &=& \,\,\, \bar\phi(x,t)  \epsilon \, ,   {\hskip 1.0cm}
   \delta \bar\phi(x,t) =  0   \, .
 \label{SUSY1}
\end{eqnarray}
  We  note that the square of this  transformation  vanishes.
   If we calculate the variation  of
 ${\bf \Sigma}$ under the transformation~(\ref{SUSY1}), we obtain
  using equation~(\ref{eq:susyaction1}):
 \begin{equation}
 \delta {\bf \Sigma} =  \bar\phi(
     {\delta\dot\phi} + \Gamma_0
 \frac{\delta^2 {\mathcal U}}  {\delta  \phi^2}  \delta  \phi)
  -  \delta{c}\, (\frac{\partial}{\partial t} + \Gamma_0
 \frac{\delta^2{\mathcal U}}  {\delta \phi^2}) \bar{c}
 -  \Gamma_0  \frac{\delta^3{\mathcal U}}  {\delta \phi^3}\delta\phi
   \, \,  c \bar{c} = - \bar\phi (\dot{ \bar{c}}  \epsilon +
   \Gamma_0
 \frac{\delta^2 {\mathcal U}}  {\delta  \phi^2} \bar{c} \epsilon)
  -  \bar\phi \epsilon (\dot{ \bar c } +  \Gamma_0
  \frac{\delta^2{\mathcal U}}  {\delta \phi^2} \bar{c}) = 0 \, .
\label{VariaSigma1}
 \end{equation}
 This expression vanishes identically because of algebraic
   anti-commutation rules. We  note  that  we do not need
 to suppose   that the potential ${\mathcal U}$ is time-independent.
\hfill\break

 $\bullet$ {\bf Invariance under   BRST2  transformation:}
 The  transformation  BRST2  mixes the different fields
   as follows~:
\begin{eqnarray}
    \delta \phi(x,t) &=&  c(x,t) \bar\epsilon \, ,   {\hskip 2.75 cm }
\delta  c(x,t)  =  0   \, ,
  \nonumber    \\
\delta \bar  \phi(x,t) &=&   \frac{\beta}{\Gamma_0}\,
\dot{c}(x,t) \bar\epsilon  \, , {\hskip 2.25 cm }
  \delta \bar{c}(x,t) =
  \left(\bar\phi(x,t) - \frac{\beta}{\Gamma_0}
 \dot\phi(x,t)\right)\bar\epsilon  \, ,
 \label{SUSY2}
\end{eqnarray}
 $\bar\epsilon$ being  a time-independent  infinitesimal Grassmann field.
  Here again,  the square of the  transformation~(\ref{SUSY2})
  vanishes.
 If we calculate the variation  of
 ${\bf \Sigma}$ under the transformation~(\ref{SUSY2}), we obtain
 \begin{equation}
 \delta {\bf \Sigma} =
  \frac{{\rm d}}{{\rm d}t} \left\{
\left( \frac{\beta}{\Gamma_0} \dot\phi - \bar\phi \right) c \right\}
  \bar\epsilon
 +\beta  \frac{\delta{\mathcal U}}{\delta \phi} \dot{c} \bar\epsilon
 + \beta  \frac{\delta^2{\mathcal U}}{\delta \phi^2} \dot\phi c \bar\epsilon
 = \frac{{\rm d}}{{\rm d}t} \left\{ \left(
    \frac{\beta}{\Gamma_0} \dot\phi +
    \beta \frac{\delta{\mathcal U}}{\delta \phi}   - \bar\phi\right) c
   \right\}\bar\epsilon
    - \beta \frac{\delta^2{\mathcal U}}{\delta \phi \partial t} c
      \bar\epsilon \, .
 \label{VariaSigma2}
 \end{equation}
   If  the potential  ${\mathcal U}$ is independent  of  time
  the last term vanishes  and
  ${\bf \Sigma}$ is invariant under the transformation~(\ref{SUSY2})
   only up to
 a  total time-derivative term  that produces   boundary  contributions to
 the total action. This does  not affect the dynamics if boundary terms vanish
  or  if the time-integral is defined from $-\infty$ to  $+\infty$.

  The  invariance  of the dynamical action ${\bf \Sigma}$
  under both   transformations
 BRST1~(\ref{SUSY1})  and BRST2~(\ref{SUSY2}) is what
  makes the time-independent dissipative Langevin equation
  supersymmetric \cite{JZJ}.
  This supersymmetric property  reflects
 the time reversal  invariance of
  Model A  in the absence of an external field and
  allows to prove the fluctuation-dissipation theorem
  \cite{gozzi, chaturvedi} as will be recalled below.

It is important to emphasize here that in the general case the use
of time reversal invariance in a path integral of the type in
Eq.(\ref{PathInt}), as mentioned above, extra precaution is
needed. Whereas the topology seen in Eq.(\ref{PathInt}) is
trivial, the general case of non-trivial topology is subtle and
time reversal invariance arguments must take this into account.
The case of non-trivial topologies was studied in details in ref.
\cite{Chernyak} where large deviations and global topological
currents were discussed in the case of an Heisenberg spin-chain
with a Wess-Zumino type term. Other work on non-trivial topologies
can be found in ref. \cite{Abanov}

\subsubsection{Ward-Takahashi  identities and the  equilibrium fluctuation-dissipation relation}
\label{subsec:FDR}

 Introducing  a four-component source
  $
  (H, \bar{H},  \bar{L}, L)$ ,
   we define  the  generating function
 \begin{eqnarray}
   Z(H, \bar{H},  \bar{L}, L)  =
  \int  {\mathcal D}\phi {\mathcal D}\bar\phi
  {\mathcal D}c {\mathcal D}\bar{c}
 \exp  \left(
  \int {\rm d}^d x      {\rm d}t
  \left( -{\bf \Sigma}(\phi ,\bar\phi,
 c ,\bar{c}) +
   \bar{H} \phi +  H \bar\phi
     + \bar{L} c  + L \bar{c}\right)
     \right)    \, .
 \label{def:Zsusy}
   \end{eqnarray}
 In the Appendix,  we rederive   the following two Ward
 identities that result from the  invariance of the action under the
 transformations~\eqref{SUSY1} and~\eqref{SUSY2}.
 The first Ward-Takahashi  identity corresponding to  invariance
  under~\eqref{SUSY1} is
 \begin{equation}
   \int {\rm d}^d x  {\rm d}t   \,  \left(
 \bar{H} \frac{\delta Z}{\delta L}  - \bar{L}  \frac{\delta Z}{\delta H}
       \right)   = 0 \, .
\label{Ward1}
\end{equation}
 The  second Ward-Takahashi  identity  corresponding to
 the transformation~\eqref{SUSY2} is  given by
 \begin{equation}
    \int {\rm d}^d x  {\rm d}t   \,  \left( \frac{\beta}{\Gamma_0}
 H \frac{{\rm d}}{{\rm d}t} \frac{\delta Z}{\delta \bar{L}}
 +{L} \left(  \frac{\delta Z}{\delta H} - \frac{\beta}{\Gamma_0}
    \frac{{\rm d}}{{\rm d}t}  \frac{\delta Z}{\delta \bar{H}} \right)
 +  \bar{H} \frac{\delta Z}{\delta \bar{L} }
       \right)   = 0 \, .
\label{Ward2}
\end{equation}

 We now  apply
  ${\delta^2}/{\delta \bar{L}(x',t')\delta  \bar{H}(x,t)}$
 to the first Ward-Takahashi
  identity~(\ref{Ward1}) and then put all the sources
 $ H, \bar{H},  L,$ and $ \bar{L}$  to 0. This leads to
 \begin{equation}
  \frac{\delta^2  Z }{\delta \bar{L}(x',t')\delta  L(x,t)} -
  \frac{\delta^2 Z }{\delta H(x',t')\delta  \bar{H}(x,t)}  = 0 \, .
 \label{id1Z}
\end{equation}
 This equation implies, using~(\ref{functderiv1}) and~(\ref{functderiv2}),
 the following identity between correlation functions:
\begin{equation}
  \langle c(x',t') \bar{c}(x,t)  \rangle =
   \langle \bar\phi(x',t') \phi(x,t)  \rangle \, .
 \label{idcorr1}
\end{equation}

     Similarly,  applying
        ${\delta^2}/{\delta \bar{H}(x',t') \delta L(x,t)}$
 to the second  Ward-Takahashi identity~(\ref{Ward2}),  and  putting  all the sources
 to zero, we obtain
\begin{equation}
 \frac{\delta^2 Z}{\delta \bar{H}(x',t') \delta H(x,t)}
  - \frac{\beta}{\Gamma_0} \,  \frac{\delta}{\delta \bar{H}(x',t')} \,\,
  \left\{ \frac{{\rm d}}{{\rm d}t}  \frac{\delta Z}{\delta \bar{H}(x,t)}
  \right\} +  \frac{\delta^2 Z} {\delta L(x,t) \delta\bar{L}(x',t')} = 0 \, .
   \label{id2Z}
\end{equation}
 This identity implies that
 \begin{equation}
  \langle \phi(x',t')  \bar\phi(x,t)  \rangle - \frac{\beta}{\Gamma_0}
   \langle \phi(x',t')   \frac{{\rm d} \phi }{{\rm d}t}(x,t)  \rangle +
  \langle  \bar{c}(x,t)  c(x',t')  \rangle = 0 \, .
  \label{idcorr2}
 \end{equation}

   In order to eliminate the correlations
 between the Grassmann variables, we combine  equation~(\ref{id1Z}) with
 equation~(\ref{id2Z}) (or equivalently
  equation~(\ref{idcorr1}) with
 equation~(\ref{idcorr2})) and use the fact that $L$ and $\bar{L}$ anti-commute
 (or  equivalently  that $c$ and $\bar{c}$ anti-commute). This leads us to:
  \begin{equation}
  \frac{\beta}{\Gamma_0}
 \langle \phi(x',t')  \dot\phi(x,t)  \rangle =
  \langle \phi(x',t')  \bar\phi(x,t)  \rangle  -
     \langle \bar\phi(x',t') \phi(x,t)  \rangle
  =   \frac{\delta}{\delta H(x,t)} \langle \phi(x',t')\rangle
   -  \frac{\delta}{\delta H(x',t')} \langle \phi(x,t)\rangle    \, .
 \label{FDTviaSUSY}
 \end{equation}
   Recalling that  $ \delta\langle\phi\rangle/\delta H$  is
 a response function, we observe that this  equation is nothing but the
  Fluctuation-Dissipation relation (see  \eqref{eq:FDT}). Usually,  the FDR is derived
 by invoking invariance under  time-reversal which implies
  detailed balance. Here, it is the  invariance  under supersymmetry that plays
 the role of  time-reversal invariance.

\subsection{Model A with a time dependent potential}

  We now study the case where the potential ${\mathcal U}[\phi(x,t),t]$
  that appears in Model A~\eqref{modelA}
 depends explicitly on time,  and show how properties
 related to supersymmetry can still be used in this nonequilibrium situation.

\subsubsection{Breakdown of the invariance for a time-dependent potential
 and its restoration  by adding the Jarzynski term}

 When the  potential  ${\mathcal U}[\phi(x,t),t]$ depends  on  time
 (by adding for example a time-dependent external field), the action
  ${\bf \Sigma}$ is no more invariant  under  supersymmetry.
  More precisely, we observed that
  invariance under~(\ref{SUSY1}) does remain
   valid even when
 ${\mathcal U}$ is a function of time and  therefore the first Ward
 identity~(\ref{Ward1}) is still satisfied. However,
   invariance under~(\ref{SUSY2})
 is broken and according to equation~(\ref{VariaSigma2}), we find
 the variation of ${\bf \Sigma}(\phi,\bar\phi,c,\bar{c})$ to be
 \begin{eqnarray}
 \delta{\bf \Sigma}(\phi,\bar\phi,c,\bar{c}) &=&
 \frac{d {\mathcal A} }{dt}  - \beta \,
\frac{\partial}{\partial t} \left(  \frac{\delta {\mathcal U}}
{\delta  \phi} \right)
  c(x,t)   \,  \bar\epsilon
 \label{varsigma}
 \end{eqnarray}
  \vskip -.3cm
 with the total derivative term
  \begin{eqnarray}
 {\mathcal A}  &=&
   \beta  \,  \Big(  \frac{\dot\phi}{\Gamma_0} +
 \frac{\delta {\mathcal U}}  {\delta  \phi}  -\frac{\bar\phi}{\beta} \Big)
   c \,  \bar\epsilon  \, . \label{eq:defA}
\end{eqnarray}
   Therefore  $\delta{\bf \Sigma}$ is not a total derivative
 and invariance under~(\ref{SUSY2})  is not true anymore for a time-dependent
 potential. In particular,
  the  second Ward identity~(\ref{Ward2}),  which was crucial
   for the proof of the Fluctuation Dissipation relation, is no more
  satisfied.

 However,  we note  that the last term in equation~\eqref{varsigma},
 which breaks the invariance
 under~(\ref{SUSY2}),  can  be rewritten as:
 \begin{equation}
   \beta \frac{\delta^2{\mathcal U}}{\delta \phi \partial t} c \,
      \bar\epsilon =  \beta
  \frac{\delta^2{\mathcal U}}{\delta \phi \partial t} \delta \phi
 = \delta\left( \beta \frac{\partial{\mathcal U}}{\partial t}\right)
 \label{contreterme}
 \end{equation}
 and  can   be interpreted as the variation
 of a function. Therefore, the   modified action
 ${\bf \Sigma_J}$,
 defined as
 \vskip -.5cm
  \begin{equation}
 {\bf \Sigma_J} = {\bf \Sigma}   + \beta \frac{\partial{\mathcal U}}{\partial t}
 \label{def:SigmaJ}
 \end{equation}
 and  obtained by   adding  the Jarzynski work~(\ref{eq:defWJ})
 to the initial action, {\it is now  invariant}  under  the
  supersymmetric BRST2-transformation ~(\ref{SUSY2})
  because its variation is given by
  a  total derivative term:
\begin{equation}
   \delta {\bf \Sigma_J}  =  \frac{d {\mathcal A} }{dt}
\label{invarianceSigmaJ}
\end{equation}
 However, we emphasize that ${\bf \Sigma_J}$ is no more  invariant
 under BRST1~\eqref{SUSY1} although ${\bf \Sigma}$  was invariant.
 We have thus restored BRST2-invariance at the expense of  BRST1.
 Therefore, in the time dependent case, neither the original
 action  ${\bf \Sigma}$ nor the  modified action  ${\bf \Sigma_J}$
 are supersymmetric \cite{JZJ}. They only exhibit partial invariances
 by either  BRST1  (in the case of  ${\bf \Sigma}$)  or BRST2
(in the case of  ${\bf \Sigma_J}$). We shall now see that
 BRST2-invariance is required to derive  nonequilibrium work identities.

 For a  time-dependent   external field $h(x,t)$,
  the   compensating  term in equation~(\ref{contreterme})  is
 given by   $- \beta  \dot{h}(x,t) \phi$.
The boundary terms at $t=\pm\infty$ are, conventionally, assumed
to vanish. Therefore, the invariance of the  dynamical action under~(\ref{SUSY2})
  breaks down when the potential is time-dependent but is restored by adding
 Jarzynski term. This observation  allows  us to use
  the    Ward-Takahashi identity~\eqref{Ward2} that results  from this
   invariance. We shall prove that this  Ward-Takahashi identity  leads
 to the nonequilibrium  work relations.

\subsubsection{Work relations from supersymmetry}

 We now show that the  invariance  of
 ${\bf \Sigma_J}$ implies the  correlator identities~(\ref{eq:deriveeniemejarz}).
 We first remark that in  the above  proofs of supersymmetric
  invariance it was noted that the boundary terms
  (a total time-derivative contribution)
  are harmless if  the integration range of the path integral is from  $-\infty$
  to $+\infty$.

  We consider an operator   ${\mathcal O}[\phi]$ that
  differs from a constant  only for $0 \le t \le  t_f $.
 Then,
  as shown in equations~\eqref{eq:pathint3.0} and \eqref{eq:pathint3bis.0},  the expectation value
 $ \langle {\mathcal O} {e}^{-\beta {\mathcal W}_J} \rangle$
 can be rewritten over an infinite time range.
  Using  Grassmann variables, we have
\begin{equation}
    \langle  {\mathcal O} {e}^{-\beta {\mathcal W}_J}  \rangle  =
    \int {\mathcal D}\phi {\mathcal D}\bar\phi
  {\mathcal D}c {\mathcal D}\bar{c} \,\,
  {\rm e}^{ - \int {\rm d}^d x  {\rm d}t
  {\bf \Sigma}(\phi,\bar\phi,c,\bar{c})} \rm{e}^{-\beta {\mathcal W}_J}
  {\mathcal O}[\phi]
  =
    \int {\mathcal D}\phi {\mathcal D}\bar\phi
  {\mathcal D}c {\mathcal D}\bar{c} \,\,
  {\rm e}^{ - \int {\rm d}^d x  {\rm d}t
  {\bf \Sigma_J}(\phi,\bar\phi,c,\bar{c})}
  {\mathcal O}[\phi]
    \, ,
  \label{eq:pathint3bis}
\end{equation}
 where all the space-time  fields  $\phi(x,t)$,  $\bar\phi(x,t)$,
 $c(x,t)$ and  $\bar{c}(x,t)$ are
  integrated   over the time interval  $ -\infty$ ~to~ $\infty$.
  In the last equation, we have combined the action   ${\bf \Sigma}$
 with the  Jarzynski work ${\mathcal W}_J$ to get the  modified action  ${\bf \Sigma_J}$,
 defined in~\eqref{def:SigmaJ}.  As shown in~\eqref{invarianceSigmaJ},
  the  modified action  ${\bf \Sigma_J}$
 is invariant under
 the transformation~\eqref{SUSY2}. Therefore,
  the  following generating function  $Z_J(H, \bar{H},  \bar{L}, L)$,
  built from the  modified action  ${\bf \Sigma_J}$,
  satisfies   the second  Ward-Takahashi identity~\eqref{Ward2}:
 \begin{eqnarray}
   Z_J(H, \bar{H},  \bar{L}, L)  =
  \int  {\mathcal D}\phi {\mathcal D}\bar\phi
  {\mathcal D}c {\mathcal D}\bar{c}
 \exp  \left(
  \int {\rm d}^d x      {\rm d}t
  \left( -{\bf \Sigma_J}(\phi ,\bar\phi,
 c ,\bar{c}) +
   \bar{H} \phi +  H \bar\phi
     + \bar{L} c  + L \bar{c}\right)
     \right)   \, .
 \label{def:ZJsusy}
\end{eqnarray}
 We  apply  the   following operator
  to  the    Ward-Takahashi identity~\eqref{Ward2}  satisfied by   $Z_J$,
  \begin{equation}
  \frac{\delta}{\delta L(x,t)}  \,
 \prod_{i=1}^n  \left(  \frac{\delta}{\delta H(x_i,t_i)}
    -\frac{\beta}{\Gamma_0}  \frac{{\rm d}}{{\rm d}t_i}
  \frac{\delta}{\delta \bar{H}(x_i,t_i) } \right)    \, ,
 \end{equation}
 and   set  the source fields  $H, \bar{H},  \bar{L}, L$ to zero. For $n=1$,
 we find
 \begin{equation}
   \left\langle
 \Big( \bar\phi(x,t) -  \frac{\beta}{\Gamma_0} \dot\phi(x,t) \Big)
 {\rm e}^{-\beta  W_J}  \right\rangle  = 0 \,
  \label{deriv1Jarz}
 \end{equation}
  More generally, for   $n \ge 1$, we have
 \begin{equation}
 \left\langle
 (\bar\phi_1 -  \frac{\beta}{\Gamma_0} \dot\phi_1)
  (\bar\phi_2 -  \frac{\beta}{\Gamma_0} \dot\phi_2) \ldots
 (\bar\phi_n -  \frac{\beta}{\Gamma_0} \dot\phi_n)
     {\rm e}^{-\beta  W_J} \right\rangle  = 0 \, ,
 \label{dJarznieme}
 \end{equation}
 where  $\phi_1 = \phi(x_1,t_1)$ etc...
 These two relations are identical to equations~(\ref{eq:deriveejarz})
  and (\ref{eq:deriveeniemejarz}),  respectively.
 In Appendix  \ref{app:converse},  we show    that  these relations
  are  equivalent to  Jarzynski's identity.
  This  concludes the proof that Jarzynski's  relation  can be obtained as
    a consequence  of a  Ward-Takahashi identity  that  itself results  from
 supersymmetric invariance.

 \section{Conclusion}
 \label{sec:conclusion}

    We have used  field-theoretic  methods  to derive nonequilibrium  work identities
    for a space-time field driven by a non-linear stochastic equation (Model A).
      We  have   obtained    a  generalization of the
     fluctuation-dissipation relation that remains valid far from equilibrium and that
    characterizes the response of a system to infinitesimal  perturbations
    around a given protocol.
   The introduction  of auxiliary fermionic fields  has allowed us to explore
   general  symmetries of the dynamical action. In particular, it is well-known
   that  the time  independent Langevin equation exhibits a hidden
   supersymmetric invariance \cite{parisi,tsvelik} that is known
   to imply  the classic fluctuation-dissipation theorem
 and Onsager's relations \cite{gozzi,chaturvedi,JZJ}.
   However, this invariance   breaks down
when  the potential varies according to a time-dependent protocol
 and drives the system out of equilibrium.
  In this work, we  have shown
 that the invariance of the effective action
 under  supersymmetric transformation is restored by adding  to
 the action  a  counter-term
 which is precisely  the  Jarzynski work ${\mathcal W}_J$.
  Furthermore, we proved  that
    the associated  supersymmetric  Ward-Takahashi  identity implies
   Jarzynski's theorem.
  Hence, supersymmetry enforces   the exactness of
 the adiabatic limit even for processes that have a finite duration
 and that can bring the system arbitrarily  far from equilibrium. In other words,
 weighing all averages with the Jarzynski  work (which amounts to modifying
 the dynamical action by adding to it the  Jarzynski  work, as in \eqref{def:SigmaJ}) restores
 one of the fundamental symmetries valid in  equilibrium. Thanks to
  this invariance,    many properties
 of the system  are effectively the same as if it were at equilibrium (although it is
 neither  in  equilibrium nor  in a stationary state).  The idea
 of considering  weighed averages, or equivalently modified path-measure
 (as was emphasized  by Jarzynski  himself in his early works \cite{jarzynskiPRE}),   allows
 to  preserve certain crucial  symmetries and  has  striking consequences in the present context.
 We believe that  a  similar arguments  should apply in many different fields: in particular,
  supersymmetry  exists   in classical
 Hamiltonian systems \cite{gozzi2} for which Jarzynski's equality
 was initially derived, and can   be  applied to prove the
 the fluctuation theorem for stochastic dynamics \cite{Kurchan}.
 Besides, the  response-field  method that we have used here
  can be extended to multi-component
    fields, to   other stochastic models with conserved order parameter  and
    also   to systems with colored  noise \cite{enpreparation}.

\acknowledgments{We  are grateful to Jean Zinn-Justin for
 many insightful conversations and remarks. K.M. thanks S. Mallick for a careful reading
 of the manuscript. M.M thanks the theory group at Saclay for its hospitality.}

 \appendix

 \section{Proof of the equivalence between equation~(\ref{eq:deriveejarz})
  and Jarzynski's relation}
  \label{app:converse}

 In this Appendix, we prove that the
   correlator identity~(\ref{eq:deriveejarz}) implies
  Jarzynski's relation~(\ref{eq:Jarzynski})  and is therefore
 equivalent to it.
   First,   we  modify   the applied  external field
  $h(x, t)$   by considering   $h(x, \alpha t)$
  for any  $\alpha > 0$. We then evaluate the average value
 $ \langle  {e}^{- \beta W_J} \rangle (\alpha)$ using
   expression~\eqref{eq:pathint3bis.0}.
 From \eqref{eq:action} and \eqref{eq:defWJ}, we observe that
  the external field  $h(x, \alpha t)$   appears only
 in the following two terms:
 $ \Gamma_0 \bar\phi(x,t) h(x,\alpha t)
   +  \beta \alpha \dot{h}(x,\alpha t)\phi(x,t) \, $
 (note that  the Jacobian term,
${\delta^2{\mathcal U}} / {\delta \phi^2}$
 does not contain $h$).
 Thus, we have
  \begin{equation}
 \frac{d \langle  {e}^{- \beta W_J} \rangle   }{d \alpha}
 =  \int {\mathcal D}\phi {\mathcal D}\bar\phi
 {\rm e}^{ -\beta {\mathcal W}_J - \int {\rm d}^d x  {\rm d}t
  \Sigma(\phi,\dot\phi,\bar\phi)}
  \int {\rm d}^d x  {\rm d}t
  \left(  \Gamma_0 t \bar\phi(x,t) \dot{h}(x,\alpha t)
      +   \beta (\dot{h}(x,\alpha t)
 + t \,  \alpha    \ddot{h}(x,\alpha t) )\phi(x,t)    \right) \, .
 \end{equation}
 Integrating by parts the last term  with  respect to time
  leads to
  \begin{eqnarray}
 \frac{d \langle {e}^{- \beta W_J} \rangle }{d \alpha}
  &=&    \int {\mathcal D}\phi {\mathcal D}\bar\phi
  { e}^{  -\beta {\mathcal W}_J - \int {\rm d}^d x  {\rm d}t
  \Sigma(\phi,\dot\phi,\bar\phi)}
  \int {\rm d}^d x  {\rm d}t
  \left(    \Gamma_0 \,  t  \bar\phi(x,t) \dot{h}(x,\alpha t)
     -   \, \beta \,  t  \dot{h}(x,\alpha t)\dot\phi(x,t)    \right) \\
   &=&  \int {\mathcal D}\phi {\mathcal D}\bar\phi
  { e}^{  -\beta {\mathcal W}_J - \int {\rm d}^d x  {\rm d}t
  \Sigma(\phi,\dot\phi,\bar\phi)}
  \int {\rm d}^d x  {\rm d}t \,  t   \dot{h}(x,\alpha t)
  \left(   \Gamma_0 \bar\phi(x,t) - \beta \dot\phi(x,t)    \right) \, .
 \end{eqnarray}
 Note that the  boundary terms that  result from  integration by parts vanish
 because we are integrating for $t \in (-\infty,+\infty)$ and because
 $\dot h = 0$ outside the  time interval $0 \le \alpha t \le t_f$.
  The last equality can be rewritten as
  \begin{equation}
  \frac{d \langle \rm{e}^{- \beta W_J} \rangle }{d \alpha} = \Gamma_0
   \int {\rm d}^d x  {\rm d}t   \,  t  \,  \dot{h}(x,\alpha t)   \left\langle
 \Big( \bar\phi(x,t) -  \frac{\beta}{\Gamma_0} \dot\phi(x,t) \Big)
 {\rm e}^{-\beta  W_J}  \right\rangle  \,  ,
 \end{equation}
 which vanishes because of  equation~(\ref{eq:deriveejarz}).
  We thus  have
   \begin{equation}
  \frac{d \langle \rm{e}^{- \beta W_J} \rangle }{d \alpha} =  0 \, .
 \end{equation}
 Hence, the  value of  $\langle \rm{e}^{- \beta W_J} \rangle$ does not
 depend on $\alpha$ and can be  evaluated
   by taking the limit    $\alpha \to 0$ which
  corresponds to an  adiabatic evolution. But then it is well known, from
 classical thermodynamics,  that
    $ W_J = -\Delta F$. This   implies
   $\langle \rm{e}^{- \beta W_J} \rangle = \exp(-\Delta F)$.

 \section{Supersymmetric formalism}
   \label{app:SUSYformalism}

 In this  appendix, we use  the superfield formalism, as explained
 in~\cite{JZJ} and~\cite{MMJZJ}, to rewrite the dynamical action and to interpret
 the invariances under the  transformations
 (\ref{SUSY1}) and (\ref{SUSY2}) in more compact and elegant language.
  In this formalism, the origin of these symmetries will appear
  more clearly. Besides, the effect of adding the Jarzynski term
 to make the time-dependent action invariant under~(\ref{SUSY2}),
 will also become more transparent.

   We introduce  two anti-commuting coordinates $\theta$ and $\bar\theta$ and
   define the superfield
\begin{equation}
{\bf \Phi}(x,t,\theta,\bar\theta)  =  \phi(x,t) + \theta \bar{c}(x,t)
 + c(x,t)\bar\theta + \theta\bar\theta \bar\phi(x,t) \, .
\label{def:superphi}
  \end{equation}
 In terms of this superfield,
 the action ${\bf \Sigma}(\phi ,\bar\phi,
 c ,\bar{c})$, defined in ~(\ref{eq:susyaction1}),
  can be written as
\begin{equation}
  {\bf \Sigma}  (\phi,\bar\phi,c,\bar{c})=
 \int d\bar\theta d\theta \,\,   {\bf \Sigma}({\bf \Phi}) \,\,\, {\rm with }
  \,\,\,   {\bf \Sigma}({\bf \Phi}) =    \Gamma_0
      \left( \bar{D}{\bf \Phi} \, D{\bf \Phi}
  + {\mathcal U}({\bf \Phi})    \right) \, ,
  \label{ActionPHI}
  \end{equation}
  where the differential operators $D$ and $\bar{D}$ are given by:
 \begin{eqnarray}
        D &=&  \frac{1}{\beta} \frac{\partial}{\partial\bar\theta} \,, \\
       \bar{D}   &=&  \frac{\partial}{\partial\theta}
  -\frac{\beta}{\Gamma_0}\bar\theta  \frac{\partial}{\partial t} \, .
\end{eqnarray}
 These two operators satisfy the anticommutation relations
 $D^2 =  \bar{D}^2 = 0$ and $\{D, \bar{D} \} = -\frac{1}{\Gamma_0}
\frac{\partial}{\partial t}\,.$

 Integration with respect to
 the  Grassmann variables is defined through the following rules \cite{JZJ}:
\begin{eqnarray}
 \int d\bar\theta d\theta \,\, 1  = 0 \, , \,\, \,\,
  \int d\bar\theta d\theta \,\, \theta  = 0 \,,  \,\, \,\,
  \int d\bar\theta d\theta \,\,  \bar\theta = 0 \, , \,\,\, \,
 \int d\bar\theta d\theta \,\,  \theta \bar\theta = 1 \, .
\end{eqnarray}
(Integration and derivation are in fact identical).

 The action $\Sigma$ is invariant under the  transformations
 (\ref{SUSY1}) and (\ref{SUSY2}) which act by mixing  the four fields
 $(\phi,\bar\phi,c,\bar{c})$. These two symmetries
 can be viewed as transformations of the superfield $\Phi$ that leave
 the super-action ${\bf \Sigma}({\bf \Phi})$  invariant.

 In the superspace formalism, the transformation~\eqref{SUSY1}
  corresponds  to   an infinitesimal translation of the $\theta$ coordinate,
   $\theta \rightarrow \theta + \epsilon$. The
  generator of this transformation is  given by
 \begin{equation}
 Q =  \frac{\partial}{\partial\theta} \, .
\label{def:Qsusy1}
\end{equation}
 Indeed, one can check that   the  superfield
 $\delta{\bf \Phi} = \epsilon Q {\bf \Phi}
 = \delta\phi(x,t) + \theta \delta\bar{c}(x,t)
 + \delta{c}(x,t)\bar\theta + \theta\bar\theta \delta\bar\phi(x,t)$
 is  given by
  $\delta{\bf \Phi} =  \epsilon \bar{c}(x,t) +
 \epsilon  \bar\theta \bar\phi(x,t) = -  \bar{c}(x,t) \epsilon
+ \bar\theta \bar\phi(x,t) \epsilon $.
 If we identify each of the   components we
 retrieve the transformation~\eqref{SUSY1}.

  Similarly,  the transformation~(\ref{SUSY2}) corresponds to
 $ \bar \theta \rightarrow \bar\theta + \bar\epsilon$ {\it and}
   $ t  \rightarrow  t + \frac{\beta}{\Gamma_0} \bar\epsilon\theta$.
  This transformation is generated  by the operator
\begin{equation}
  \bar{Q} =  \frac{\partial}{\partial\bar\theta}
   + \frac{\beta}{\Gamma_0} \, \theta \, \frac{\partial}{\partial t} \,.
  \label{def:Qbarsusy2}
 \end{equation}

 The  operators $Q$ and $\bar{Q}$  that generate the supersymmetry
 transformations   anticommute with  $D$ and $ \bar{D}$. Besides, they
 satisfy the anticommutation relations
 $Q^2 =  \bar{Q}^2 = 0$ and $\{Q, \bar{Q} \} = \frac{\beta}{\Gamma_0}
\frac{\partial}{\partial t}\,.$
 When the potential  ${\mathcal U}$ does not depend on time,
 the action ${\bf \Sigma}({\bf \Phi})$ is symmetric  under  $Q$,
 and is invariant under  $\bar{Q}$  upto a total derivative. This fact
 was checked  in equations~(\ref{VariaSigma1},\ref{VariaSigma2})
  and  can be  verified again  using
 the supersymmetry formalism. If  ${\mathcal U}$ depends explicitly on time
 the action ${\bf \Sigma}({\bf \Phi})$ is not invariant anymore under
 $\bar{Q}$. However, by adding to it the Jarzynski term~(\ref{eq:defWJ}),
 we obtain the modified action
  ${\bf \Sigma_J}$, defined in equation~\eqref{def:SigmaJ},
 which is invariant under   $\bar{Q}$.  This property is
 manifest in the supersymmetric formalism in which  the modified action
 is written as
  \begin{equation}
 {\bf \Sigma_J} =  \Gamma_0
      \left( \bar{D}{\bf \Phi} \, D{\bf \Phi}
  + {\mathcal U}({\bf \Phi}, t + \frac{\beta}{\Gamma_0} \theta \bar{\theta})
    \right) \, .
  \end{equation}

 \section{Supersymmetric Ward-Takahashi Identities}
   \label{demoward}

 When the invariances under the  transformations
 generated by the operators~\eqref{def:Qsusy1} and~\eqref{def:Qbarsusy2}
 are implemented in   the generating function   $Z(H, \bar{H},  \bar{L}, L)$
 defined in~\eqref{def:Zsusy},
  the   Ward-Takahashi  Identities \eqref{Ward1} and \eqref{Ward2} are
obtained.  We follow  closely the method of \cite{chaturvedi} to derive these
 identities.
 In order to calculate correlation functions it is helpful
 to rewrite  the  sources  as a superfield ${\bf J}$,  defined as:
 \begin{equation}
 {\bf J} = H + \theta\bar{L} + \bar\theta L +  \theta\bar\theta \bar{H}
 \end{equation}
  where $L(x,t)$ and $\bar{L}(x,t)$ are Grassmann  fields.  We  thus have
\begin{equation}
  \int d\bar\theta  d\theta {\bf J}(x,t,\theta,\bar\theta)
    {\bf \Phi}(x,t,\theta,\bar\theta) = \bar{H} \phi +  H \bar\phi
     + \bar{L} c  + L \bar{c}  \, .
\end{equation}
  We note from this expression
 that  $H$ plays the role of an applied external `magnetic' field.
 In this formalism, the generating function  $Z(H, \bar{H},  \bar{L}, L)$
 becomes
 \begin{equation}
   {Z}({\bf J})  =   \int {\mathcal D}{\bf \Phi}
  e^{\int {\rm d}^d x  {\rm d}t  d\bar\theta d\theta
 (-  {\bf \Sigma}({\bf \Phi})  +  {\bf J}{\bf \Phi} ) } .
 \label{def:Zsuperfield}
       \end{equation}
 (In the sequel, the
 integration element ${\rm d}^d x  {\rm d}t  d\bar\theta d\theta$
 will be omitted in general.)

  In order to derive the first Ward-Takahashi identity, we proceed as follows.
   In the  functional integral~(\ref{def:Zsuperfield}),  we make
  the change of variable ${\bf \Phi}  \rightarrow   {\bf \tilde \Phi}$
  with $ {\bf \Phi} = {\bf\tilde\Phi} + \epsilon Q {\bf\tilde \Phi} $,
 where $Q$, defined in~\eqref{def:Qsusy1}
 is the infinitesimal  generator of the transformation~\eqref{SUSY1}
 corresponding to $\theta$ translations. Taking into account that the Jacobian is 1, we obtain
  \begin{equation}
  { Z}({\bf J})  =   \int {\mathcal D}{\bf \Phi}
  {e}^{\int  -  {\bf \Sigma}({\bf \Phi}) +  {\bf J}{\bf \Phi}}
 =  \int {\mathcal D}{\bf \tilde \Phi}
  {e}^{\int  -  {\bf \Sigma}({\bf\tilde\Phi} + \epsilon Q {\bf\tilde \Phi})
   +  {\bf J}({\bf\tilde\Phi} + \epsilon Q {\bf\tilde \Phi}) }
 =  \int {\mathcal D}{\bf  \Phi}
  {e}^{\int  -  {\bf \Sigma}({\bf \Phi} + \epsilon Q {\bf  \Phi})
   +  {\bf J}({\bf \Phi} + \epsilon Q {\bf  \Phi}) }  \, .
 \label{chgtvarZ}
 \end{equation}
 (The last equality   results simply from the fact that ${\bf\tilde \Phi}$
 is a dummy variable.)
 The fact that the action is invariant means precisely that
  $ \int  d\bar\theta d\theta {\bf \Sigma}({\bf \Phi}) =
 \int  d\bar\theta d\theta {\bf \Sigma}({\bf \Phi} + \epsilon Q {\bf  \Phi}).$
 Therefore, we deduce that
  \begin{equation}
  {Z}({\bf J})  =   \int {\mathcal D}{\bf \Phi}
  {e}^{\int  -  {\bf \Sigma} +  {\bf J}{\bf \Phi}}
 =   \int {\mathcal D}{\bf  \Phi}
 {e}^{\int  -  {\bf \Sigma} +  {\bf J}{\bf \Phi} +
   {\bf J} \epsilon Q {\bf  \Phi} } =
   \int {\mathcal D}{\bf  \Phi}
 {e}^{\int  -  {\bf \Sigma} +  {\bf J}{\bf \Phi}} ( 1 + \epsilon \int
{\bf J}  Q {\bf  \Phi} ) \, .
 \label{vardeZ}
  \end{equation}
 This equation being true for any value of $\epsilon$ we conclude that
 \begin{equation}
    \int {\mathcal D}{\bf \Phi} \left\{
    \int {\rm d}^d x  {\rm d}t  d\bar\theta d\theta  \,
    {\bf J}  Q {\bf\Phi}  \right\}
    {e}^{\int  -  {\bf \Sigma} +  {\bf J}{\bf \Phi}}    = 0 \, .
\label{PreWard1}
\end{equation}
 We now calculate explicitly the value of ${\bf J}.Q{\bf\Phi}$
 and substitute it in equation~(\ref{PreWard1}),
 \begin{equation}
    \int {\mathcal D}{\bf \Phi} \left\{
    \int {\rm d}^d x  {\rm d}t   \,
         \bar{H}(x, t) \bar{c}(x,t) - \bar{L}(x,t) \bar\phi(x,t)  \right\}
    {e}^{\int  -  {\bf \Sigma} +  {\bf J}{\bf \Phi}}    = 0 \, .
\label{semiWard1}
\end{equation}
  By differentiating  the generating function $Z({\bf J})$ with respect to
 $\bar{H}$,  we obtain
  \begin{equation}
        \frac{\delta Z({\bf J})}{\delta \bar{H}(x_a, t_a)}
  =  \int {\mathcal D}{\bf \Phi} \, \phi(x_a, t_a)
     exp^{\int  -  {\bf \Sigma} +  {\bf J}{\bf \Phi}} \, .
 \label{functderiv1}
  \end{equation}
  Similarly, we have
  \begin{equation}
\frac{\delta Z}{\delta L} \rightarrow \bar{c} \, ,
  {\hskip 1cm}  \frac{\delta Z}{\delta \bar{L} } \rightarrow {c} \, ,
   {\hskip 0.7cm} \hbox{ and } \,\,\,
    \frac{\delta Z}{\delta H} \rightarrow \bar\phi \, .
  \label{functderiv2}
   \end{equation}
 Substituting these relations in equation~(\ref{semiWard1})  allows
 us to derive the first Ward-Takahashi identity~\eqref{Ward1}:
 \begin{equation}
   \int {\rm d}^d x  {\rm d}t   \,  \left(
 \bar{H} \frac{\delta Z}{\delta L}  - \bar{L}  \frac{\delta Z}{\delta H}
       \right)   = 0 \, .
\end{equation}

  For  the second invariance under the
  BRST2  transformation~\eqref{SUSY2}, we
 use  the   infinitesimal  generator  $\bar{Q}$, defined
  in~\eqref{def:Qbarsusy2}.  After similar
  calculations, we find
 \begin{equation}
    \int {\mathcal D}{\bf \Phi} \left\{
    \int {\rm d}^d x  {\rm d}t  d\bar\theta d\theta
    {\bf J} \bar{Q}  {\bf  \Phi}  \right\}
    {e}^{\int  -  {\bf \Sigma} +  {\bf J}{\bf \Phi}} = 0    \, .
 \label{PreWard2}
\end{equation}
  After calculating explicitly  ${\bf J} \bar{Q}{\bf\Phi}$, we obtain
\begin{equation}
    \int {\mathcal D}{\bf \Phi} \left\{
    \int {\rm d}^d x  {\rm d}t   \, \frac{\beta}{\Gamma_0}
{H}(x, t) \dot{c}(x,t)  + {L}(x,t) ( \bar\phi(x,t) -
 \frac{\beta}{\Gamma_0} \dot\phi(x,t) ) +   \bar{H}(x, t) {c}(x,t) \right\}
   {e}^{\int  -  {\bf \Sigma} +  {\bf J}{\bf \Phi}}    = 0 \, .
\label{semiWard2}
\end{equation}
 Expressing the fields in this equation as  functional derivatives
 of the generating function $Z$, leads us to  the
  second Ward-Takahashi  identity~\eqref{Ward2}:
 \begin{equation}
    \int {\rm d}^d x  {\rm d}t   \,  \left( \frac{\beta}{\Gamma_0}
 H \frac{{\rm d}}{{\rm d}t} \frac{\delta Z}{\delta \bar{L}}
 +{L} \left(  \frac{\delta Z}{\delta H} - \frac{\beta}{\Gamma_0}
    \frac{{\rm d}}{{\rm d}t}  \frac{\delta Z}{\delta \bar{H}} \right)
 +  \bar{H} \frac{\delta Z}{\delta \bar{L} }
       \right)   = 0 \, .
\end{equation}

\end{document}